\documentclass[prl,twocolumn,showpacs,amsmath,amssymb,superscriptaddress]{revtex4-2}
\usepackage{dcolumn}
\usepackage{bm,graphicx}
\usepackage{etoolbox}
\usepackage{url}
\usepackage{amsmath,amsfonts,amsthm}

\usepackage[colorlinks]{hyperref}
\usepackage{breakurl}
\usepackage{color}
\usepackage{bbold}
\usepackage [latin1]{inputenc}

\DeclareMathAlphabet\mathbfcal{OMS}{cmsy}{b}{n}

\begin{document}

\title{Lorentz-boost-driven magneto-optics in a Dirac nodal-line semimetal}

\author{J.~Wyzula}
\affiliation{LNCMI-EMFL, CNRS UPR3228, Univ. Grenoble Alpes, Univ. Toulouse, Univ. Toulouse 3, INSA-T,  Grenoble and Toulouse, France}

\author{X.~Lu}
\affiliation{Laboratoire de Physique des Solides, Universit\'e Paris Saclay, CNRS UMR 8502, F-91405 Orsay Cedex, France}

\author{D.~Santos-Cottin}
\affiliation{Department of Physics, University of Fribourg, Chemin du Mus\'ee 3, 1700 Fribourg, Switzerland}

\author{D. K. Mukherjee}
\affiliation{Laboratoire de Physique des Solides, Universit\'e Paris Saclay, CNRS UMR 8502, F-91405 Orsay Cedex, France}
\affiliation{Department of Physics, Indiana University, Bloomington, Indiana 47405, USA}

\author{I.~Mohelsk\'y}
\affiliation{LNCMI-EMFL, CNRS UPR3228, Univ. Grenoble Alpes, Univ. Toulouse, Univ. Toulouse 3, INSA-T,  Grenoble and Toulouse, France}

\author{F. Le Mardel\'e}
\affiliation{Department of Physics, University of Fribourg, Chemin du Mus\'ee 3, 1700 Fribourg, Switzerland}

\author{J.~Nov\'{a}k} 
\affiliation{Department of Condensed Matter Physics, Masaryk University, Kotl\'a\v{r}sk\'a 2, 611 37 Brno, Czech Republic}

\author{M.~Novak} 
\affiliation{Department of Physics, Faculty of Science, University of Zagreb, 10000 Zagreb, Croatia}

\author{R.~Sankar}
\affiliation{Institute of Physics, Academia Sinica, Nankang, 11529 Taipei, Taiwan}

\author{Y.~Krupko}
\affiliation{LNCMI-EMFL, CNRS UPR3228, Univ. Grenoble Alpes, Univ. Toulouse, Univ. Toulouse 3, INSA-T,  Grenoble and Toulouse, France}
\affiliation{Institut d'Electronique et des Systemes, UMR CNRS 5214, Universit\'e de Montpellier, 34000, Montpellier, France}

\author{B.~A.~Piot}
\affiliation{LNCMI-EMFL, CNRS UPR3228, Univ. Grenoble Alpes, Univ. Toulouse, Univ. Toulouse 3, INSA-T,  Grenoble and Toulouse, France}

\author{W.-L.~Lee}
\affiliation{Institute of Physics, Academia Sinica, Nankang, 11529 Taipei, Taiwan}

\author{A.~Akrap}
\affiliation{Department of Physics, University of Fribourg, Chemin du Mus\'ee 3, 1700 Fribourg, Switzerland}

\author{M.~Potemski}
\affiliation{LNCMI-EMFL, CNRS UPR3228, Univ. Grenoble Alpes, Univ. Toulouse, Univ. Toulouse 3, INSA-T,  Grenoble and Toulouse, France}
\affiliation{Institute of Experimental Physics, Faculty of Physics, University of Warsaw, ul. Pasteura 5, 02-093 Warszawa, Poland}

\author{M.~O.~Goerbig}
\affiliation{Laboratoire de Physique des Solides, Universit\'e Paris Saclay, CNRS UMR 8502, F-91405 Orsay Cedex, France}

\author{M.~Orlita}
\email[]{milan.orlita@lncmi.cnrs.fr}
\affiliation{LNCMI-EMFL, CNRS UPR3228, Univ. Grenoble Alpes, Univ. Toulouse, Univ. Toulouse 3, INSA-T,  Grenoble and Toulouse, France}
\affiliation{Charles University, Faculty of Mathematics and Physics, Institute of Physics, Ke Karlovu 5, 121 16 Prague 2, Czech Republic}


\begin{abstract}
Optical response of crystalline solids is to a large extent driven by excitations that promote electrons among individual bands. This allows one to apply optical and magneto-optical methods to determine experimentally the energy band gap -- a fundamental property crucial to our understanding of any solid -- with a great precision. Here we show that such conventional methods, applied with great success to many materials in the past, do not work in topological Dirac semimetals with a dispersive nodal line. There, the optically deduced band gap depends on how the magnetic field is oriented with respect to the crystal axes. Such highly unusual behaviour is explained in terms of band-gap renormalization driven by Lorentz boosts which results from the Lorentz-covariant form of the Dirac Hamiltonian relevant for the nodal line at low energies.
\end{abstract}

\maketitle

\vspace{3mm}


\section{Introduction}

Ever since the advent of graphene and topological materials, relativistic physics has become an integral part of condensed-matter sciences~\cite{BansilRMP16,ChiuRMP16,ArtmitageRMP18}. While emergent, it is important to stress that this type of relativity is pertinent beyond the dispersion of the low-energy excitations in many different solids. Klein tunneling~\cite{KleinZfP29,KatsnelsonNP06,YoungNP09} and the chiral anomaly~\cite{NielsenPLB83,SonPRB13,XiongScience15,HuangPRX15,BurkovScience15,ReisNJP16,JiaNatureMater16,OngNRP21} represent well-known examples. One of the salient aspects of relativity is the particular dependence of energy on the frame of reference: for a particle of mass $m$ moving at a speed $u$ lower than the speed of light $c$, a Lorentz boost to the comoving frame of reference changes the particle's energy from $E$ to $E/\gamma=mc^2$, in terms of the Lorentz factor $\gamma=1/\sqrt{1-\beta^2}$ and the rapidity $\beta=u/c$. A natural question that one may now ask is the following:
can one observe this relativistic renormalization equally in topological materials governed by the Dirac Hamiltonian or a variant of it, upon the replacement of $c$ by a characteristic velocity $v$?\\
\vspace{1mm}

While the effects of Lorentz covariance have been theoretically studied, namely in the framework of systems with tilted conical bands, both in two~\cite{KatayamaJPSJ04,GoerbigEPL09,SariPRB15} and three~\cite{SoluyanovNature15,TchoumakovPRL16,ZhangJAP21} dimensions, an experimental verification is yet lacking. The key finding of this paper is that Lorentz boosts have important experimental consequences in Dirac materials. Most notably and unexpectedly, we find that the optical band gap extracted from magneto-optical measurements in the dispersive nodal-line Dirac semimetal niobium diarsenide (NbAs$_2$) depends on the orientation of the explored crystal. As we show below, this orientation defines a particular Lorentz boost, by fixing the angle between the applied magnetic field and the direction of the spectroscopically relevant part of the nodal line.\\
\vspace{1mm}

\begin{figure*}[t]
\begin{center}
\includegraphics[width=0.96\textwidth]{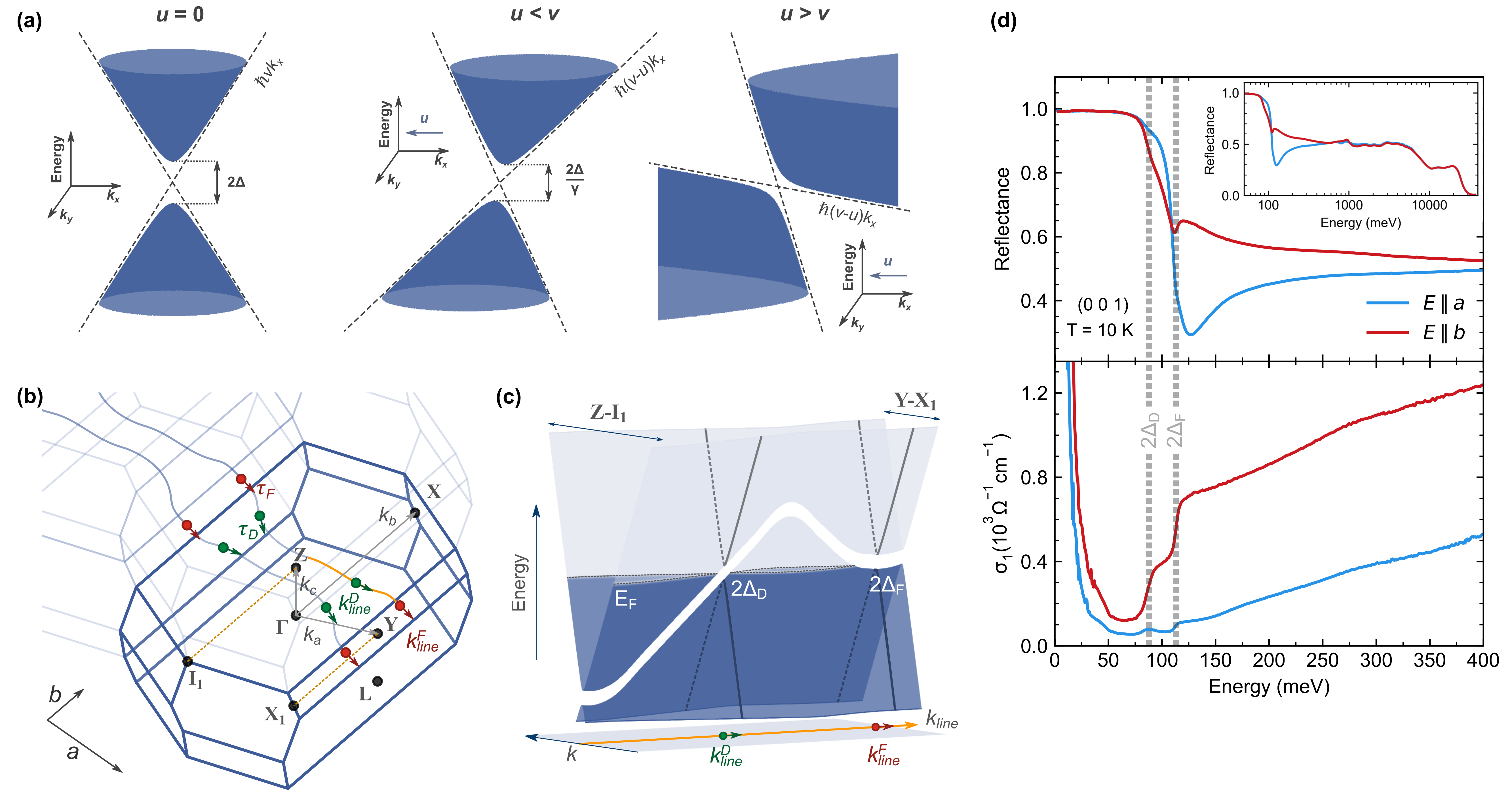}
\end{center}
\caption{ (a) Hyperbolic dispersion of massive Dirac electrons with an additional tilt: $E(\mathbf{k}) = -\hbar u k_x \pm \sqrt{\Delta^2+\hbar^2v^2k^2}$.
(b) BZ of NbAs$_2$ with selected high-symmetry points. The blue curved lines show approximate positions of two nodal lines~\cite{ShaoPNAS19}. The crossings with the Fermi level are marked by green and red full circles, the arrows indicate local nodal-line directions, (c)  Schematic view of the nodal line dispersion along the direction of propagation in a half of the BZ, highlighted using yellow color in panel (b). The band gaps at the crossing points with the Fermi energy are $2\Delta_D$ and $2\Delta_F$, for the dispersive and (approximately) flat parts of the nodal line, respectively. (d) Infrared reflectivity, and the extracted optical conductivity, measured on (001)-oriented facet of NbAs$_2$, using light polarized along the $a$ and $b$ axes.}
\label{fig:intro-panel}
\end{figure*}

To appreciate the link between magnetic-field and Lorentz boosts in tilted cones, let us first consider a 2D conical band, characterized by a gap 2$\Delta$ and an asymptotic velocity $v$, which is tilted by an additional velocity parameter $\mathbf{u}$ (in  \textbf{Figure}~\ref{fig:intro-panel}a, $\mathbf{u}\,\|\, \mathbf{\hat{x}}$). Such a system is described by the following variant of a 2D massive Dirac Hamiltonian:
\begin{equation}
\label{eq:ham2Dtilt}
    \hat{H}_\text{2D} = \hbar \mathbf{u}\cdot\mathbf{k}\,\mathbb{1}+\left[
    \begin{array}{cc}
        \Delta & \hbar v (k_x - ik_y) \\
        \hbar v (k_x+i k_y) & -\Delta
    \end{array}
    \right].
\end{equation}
In an out-of-plane magnetic field, \emph{i.e.}, with $\mathbf{B}$ applied perpendicular to the $x$-$y$ plane, and thus also to $\mathbf{u}$, the tilt can be formally viewed as a drift velocity of electrons in the crossed magnetic and effective electric fields, ${\cal E}=uB$. The drift is in the direction perpendicular to both the tilt and the magnetic field.\\
\vspace{1mm}

In this specific case, the problem of the electron motion in a tilted cone becomes mathematically equivalent to the dynamics of a relativistic charge carrier in the crossed electric and magnetic fields~\cite{LukosePRL07,GoerbigEPL09,TchoumakovPRL16}. This motion is therefore governed by fully Lorentz-covariant Dirac and Maxwell equations. This covariant formulation, and thus the use of Lorentz transformations, allows us to calculate the energy spectrum in a reference frame where the (effective) electric field vanishes, meaning $u=0$~\cite{GoerbigEPL09,SariPRB15,LukosePRL07}. A similar relativistic-like approach has been invoked in the past, in order to understand behaviour of narrow-gap semiconductors~\cite{AronovJETP67,Zawadzki1985,Zawadzki1986} in real crossed electric and magnetic fields.\\
\vspace{1mm}

The impact of the tilt $\mathbf{u}$ --  or in the sense of reasoning above, the impact of the Lorentz boost -- on the Landau quatization is profound~\cite{GoerbigEPL09,TchoumakovPRL16}. We obtain the Landau level (LL) spectrum that is typical of 2D massive Dirac electrons, but whose energy band gap and velocity parameter are renormalized by the Lorentz factor, $\gamma=1/\sqrt{1-u^2/v^2}$:
\begin{equation}
\label{LL_spectrum}
E_n=\pm\sqrt{(\Delta/\gamma)^2+ 2eBn\hbar v^2/\gamma^3},\quad n=0,1,2\ldots
\end{equation}
 For large tilts, $u \geq v$, the spectrum collapses and marks a transition between
regimes referred to as magnetic and electric~\cite{TchoumakovPRL16,TakeuchiPRB02}. In the semi-classical picture, this crossover corresponds to a transition from closed towards open cyclotron orbits in type-I and type-II conical bands (Figure~\ref{fig:intro-panel}a).

\section{Optical and magneto-optical response of NbAs$_2$}

Let us now discuss how a Lorentz boost allows us to understand the magneto-optical response of NbAs$_2$ which is a nearly compensated 3D nodal-line semimetal~\cite{XuPRB16} with a monoclinic crystal lattice and the space group C2/m~\cite{BenschAC95}. Theory and experimental studies performed so far~\cite{ShaoPNAS19,XuPRB16} indicate the existence of two nodal lines in this material. Following a strict definition, these are 1D objects (curves) in momentum space, $\mathbf{k}_{\mathrm{line}}$, along which the gap vanishes. Nevertheless, due to spin-orbit interaction, the gap does not close completely in NbAs$_2$. Instead, it only reaches a local minimum, similar to all other materials which are in current literature referred to as nodal-line or nodal-loop semimetals~\cite{ShaoNP2020,SchoopNC16,TakanePRB16}. In NbAs$_2$, the nodal lines are open (periodically penetrating the Brillouin zone boundaries), they propagate approximately along the $a$ crystallographic axis, and they are located symmetrically with respect to the $\Gamma$-Y-Z mirror plane  (Figure~\ref{fig:intro-panel}b).
\vspace{1mm}

The low-energy electronic excitations around the nodal line can be described using a model for 2D massive Dirac electrons in the plane perpendicular to the local nodal-line direction $\boldsymbol \tau$ (left panel of Figure~\ref{fig:intro-panel}a). The corresponding velocity parameter $v$ and the spin-orbit gap $2\Delta$ vary smoothly along the line. Essentially for this work, the nodal lines in NbAs$_{2}$ disperse with momentum and each approaches the Fermi energy four times within the Brillouin zone (Figure~\ref{fig:intro-panel}b,c). The crossings come in pairs of two different types. One of them is associated with a dispersive and the other one with a flat part~\cite{ShaoPNAS19}. They are located at $\mathbf{k}^D_{\mathrm{line}}$ and $\mathbf{k}^F_{\mathrm{line}}$, respectively, and characterized by the local directions $\boldsymbol\tau_D$ and $\boldsymbol\tau_F$ (Figure~\ref{fig:intro-panel}b,c).
\vspace{1mm}

\begin{figure*}[t]
\begin{center}
\includegraphics[width=0.9\textwidth]{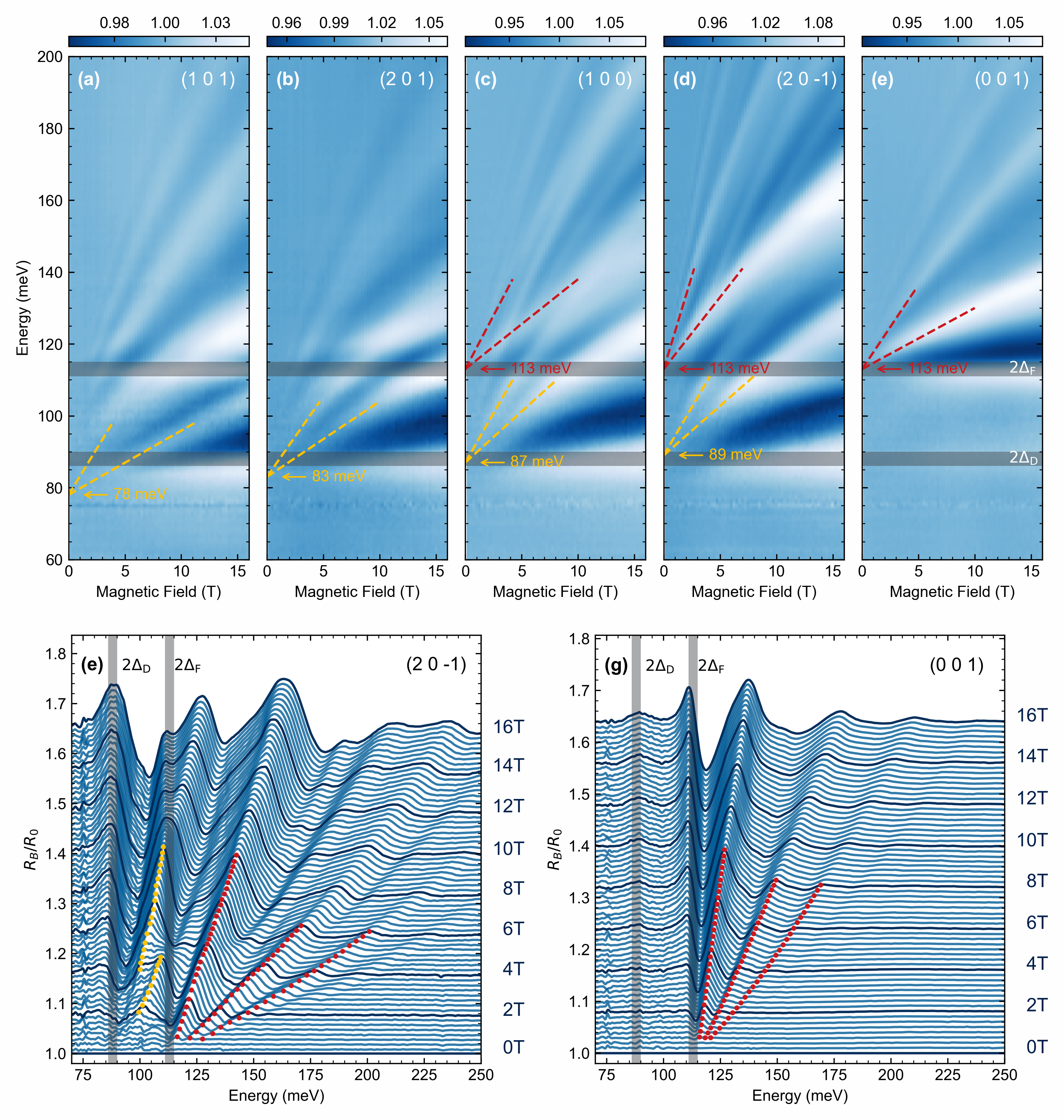}
    \end{center}
\caption{Panels a-e: False color-plots of relative magneto-reflectivity of NbAs$_2$, $R_B/R_0$, in the magnetic field applied
along five different reciprocal space directions: (101), (201), (100), (20$\bar{1}$) and (001), which make angles 62$^{\circ}$, 51$^{\circ}$, 29$^{\circ}$, 7$^{\circ}$ and 90$^{\circ}$ with the $a$ axis, respectively. The yellow and red values indicate the apparent optical band gaps, $2\Delta_D^{\mathrm{eff}}$ and $2\Delta_F$, deduced using a linear zero-field extrapolation of inter-LL resonances belonging to the lower and upper set, respectively (dashed lines).
Panels f and g: stack-plots of relative magneto-reflectivity spectra, $R_B/R_0$, for selected values of the magnetic field
collected on the (20$\bar{1}$) and (001) facets, respectively. The yellow and red dots show $R_B/R_0$ maxima belonging to transitions in the lower and upper set, respectively. The horizontal and vertical gray bars show positions of the two steps in the onset of interband absorption at $2\Delta_D$ and $2\Delta_F$ at $B=0$ (Figure~\ref{fig:intro-panel}d).}
\label{Fig:color-maps}
\end{figure*}

In the absence of a magnetic field, the case we examine first,
the optical response of NbAs$_2$ at low photon energies comprises direct signatures of nodal lines~\cite{ShaoPNAS19,MelePNAS19,SM}. Two steps are clearly visible at the onset of interband absorption in the optical conductivity (Figure~\ref{fig:intro-panel}d down). They correspond to the local band gaps: $2\Delta_D=(88\pm2)$~meV and $2\Delta_F=(113\pm2)$~meV, at the dispersive and flat parts, respectively.
Above this onset, the optical conductivity increases linearly in $\omega$. While such behavior resembles systems with 3D conical bands~\cite{TimuskPRB13,OrlitaNP14,NeubauerPRB16}, in NbAs$_2$, it is due to the occupation  effect (Pauli-blocking) along the dispersive part of the nodal line~\cite{ShaoPNAS19,MelePNAS19,Proninppsb21,Santos-CottinPRB21}. The anisotropy of the optical response (Figure~\ref{fig:intro-panel}d) reflects the orientation of the nodal lines, implying significantly higher Drude-type and interband absorption strength for the
radiation polarized along the $b$ axis which is approximately perpendicular to $\boldsymbol\tau$~\cite{ShaoPNAS19}. Strong anisotropy, implying notably elongated Fermi surfaces, was also observed in magneto-transport experiments~\cite{SM,ShenPRB16,WangPRB16,PeramaiyanSR2018}.
\vspace{1mm}

To explore the magneto-optical response, we have identified a number of crystallographic facets on several NbAs$_2$ monocrystals using the standard $x$-ray technique~\cite{SM}. Then, a series of infrared reflectivity measurements has been carried out, using non-polarized radiation and the Faraday configuration. The magnetic field was applied perpendicular to chosen crystallographic planes.  In this way, we probed electrons undergoing cyclotron motion in crystallographic planes with various orientations with respect to the local nodal-line directions $\boldsymbol\tau_D$ and $\boldsymbol\tau_F$.
To keep the geometry as simple as possible, we selected facets with a zero middle index ($n\,0\,m$).
Thus the vector $\mathbf{B}$ always lied in the mirror plane of the NbAs$_2$ crystals and formed identical angles $\theta_F$ and $\theta_D$ with the local directions, $\boldsymbol\tau_{F}$ and $\boldsymbol\tau_D$, of the two mirror-symmetric nodal lines at the flat and dispersive parts, respectively, where the Fermi level is crossed.
\vspace{1mm}

The relative magneto-reflectivity of NbAs$_2$, $R_B/R_0$, measured
with $\mathbf{B}$ applied perpendicular to the (101), (201), (100), (20$\bar{1}$) and (001) crystallographic planes, is presented in~\textbf{Figure}~\ref{Fig:color-maps}a-e
in a form of false-color plots. For illustration, we also append two stack-plots of selected reflectivity spectra collected on the (20$\bar{1}$) and (001) facets in~Figures~\ref{Fig:color-maps}f and g, respectively.
Data measured on other facets are shown in Supplementary Materials~\cite{SM}.
The observed response contains a series of well-defined resonances with a weakly sublinear dependence on $B$ that can be directly associated with interband inter-Landau-level (inter-LL) excitations.
The observed response -- in position, spacing and relative/absolute intensity of inter-LL transitions -- strongly varies with the explored facet. Across the board, this response includes two characteristic sets of interband inter-LL excitations: (i) the upper set, with transitions that always extrapolate to the energy of $2\Delta_F$ in the zero-field limit and are thus clearly associated with the flat crossing of the nodal line and (ii) the lower set of transitions that extrapolate, depending on the particular facet, to the energy equal to, or lower than $2\Delta_D$ (see $B=0$ extrapolations using yellow dashed lines in Figure~\ref{Fig:color-maps}).
For some facets, only one of these two sets is clearly manifested in the data.
\vspace{1mm}

This observation -- an orientation-dependent gap -- is the main finding of our work and it ventures beyond the common knowledge in LL spectroscopy of solids~\cite{Landwehr2012}. In conventional materials, the slope $d(\hbar\omega)/dB$, and the spacing of inter-LL excitations may depend on the orientation of the crystal with respect to the magnetic field, as well as on the  experimental configuration ({\emph e.g.}, Faraday versus Voigt). Nevertheless, the apparent band gap obtained from the zero-field extrapolation of interband inter-LL excitations is widely used as an unambiguous estimate of the separation between electronic bands. Quite unexpectedly, such an apparently evident approach fails in our case here.

\section{Theoretical model}

To show that this orientation-dependent gap is a signature of the relativistic properties in the present material, let us consider the following minimal Hamiltonian for a dispersive nodal line:
\begin{equation}\label{H0}
\hat{H}  =  \hbar\,w\,q_{\mathrm{line}} \mathbb{1} + \left[ \begin{matrix} \Delta  &\hbar v (q_x-iq_y) \\ \hbar v (q_x+iq_y)&-\Delta \end{matrix} \right],
\end{equation}
where the wave vector $\mathbf{q}=(q_x,q_y,q_{\mathrm{line}})=\mathbf{k}-\mathbf{k}_{\mathrm{line}}$ is defined with respect to any point $\mathbf{k}_\mathrm{line}$ on the nodal line. The velocity parameter $w$ describes the slope of the dispersing nodal line. In the plane $(q_x,q_y)$ perpendicular to the local direction $\boldsymbol \tau$, the Hamiltonian (\ref{H0}) becomes that of a 2D massive Dirac electron, with the gap $2\Delta$ and the asymptotic velocity $v$.
In contrast, when the bands are cut in a plane that is not perpendicular to $\boldsymbol \tau$, the corresponding dispersion shows the tilt described by the Hamiltonian (\ref{eq:ham2Dtilt}) and schematically shown in Figure~\ref{fig:intro-panel}a.
\vspace{1mm}

When a magnetic field is applied, making an angle $\theta$ with the local direction $\boldsymbol\tau$, it is the conventional Lorentz force and the particular profile of the dispersion which govern the motion of electrons in the plane perpendicular to $\mathbf{B}$. Thus, for non-zero angles $\theta$, we study the magneto-optical response of electrons in tilted anisotropic conical bands, for details see Supplementary Materials~\cite{SM}.
The corresponding LL spectrum then gets the Lorentz-boost-renormalized form of Equation~\ref{LL_spectrum}, amended by the dispersive term $\hbar wq_B/\cos\theta$, where $q_B$ is the wave vector along the applied magnetic field. The rapidity determining the Lorentz factor is calculated as the ratio of the tilt and asymptotic velocities, $w\sin\theta$ and $v\cos\theta$, respectively, so that $\beta=(w/v)\tan\theta$~\cite{SM}. Hence, depending on the orientation of the magnetic field with respect to the nodal line, we expect a pseudo-relativistic decrease of the band gap and the velocity parameter:
\begin{equation}\label{gapRenorm}
    2\Delta\rightarrow 2\Delta^\text{eff}=\frac{2\Delta}{\gamma}~~\text{and}~~ v\rightarrow v^\text{eff}=\frac{v\sqrt{\cos\theta}}{\gamma^{3/2}},
\end{equation}
in terms of the effective Lorentz factor:
\begin{equation}\label{eq:gammaeff}
    \gamma= \frac{1}{\sqrt{1-\frac{w^2}{v^2}\tan^2\theta}}.
\end{equation}
For angles exceeding the critical value of $\tan^{-1}(v/w)$, the quantization into LLs is expected to collapse in a way analogous to overtilted 3D conical bands~\cite{TchoumakovPRL16}. Compared to 2D Dirac systems,
Equation~(\ref{eq:gammaeff}) clearly shows that the Lorentz factor
$\gamma$ is now tunable by the angle $\theta$. This allows us to continuously monitor, using the angle $\theta$, the band gap renormalization until the extinction of discrete LLs (see Figure~\ref{Fig:color-maps}).

\section{Data analysis and discussion}

To analyze our experimental data quantitatively in view of the above theoretical picture, we focus on the lowest observed line in both sets and assign it to the inter-LL excitation $0 \leftrightarrow 1$. Even though the  pseudo-relativistic renormalization may profoundly alter the selection rules~\cite{SM,SariPRB15}, this transition is theoretically~\cite{SM} expected to remain strong for any $\beta<1$ and its energy reads:
\begin{equation}
   \hbar\omega_{0\leftrightarrow 1}=\Delta^{\mathrm{eff}}+\sqrt{(\Delta^{\mathrm{eff}})^2+2e\hbar B (v^{\mathrm{eff}})^2}.
\end{equation}
This expression, as well as Equations~(\ref{gapRenorm}) and (\ref{eq:gammaeff}) are valid for both optically active parts of the nodal line. In the flat part around $\mathbf{k}_\text{line}^F$, we have $w=0$, i.e., $\gamma=1$, so that one does not expect any facet dependence of the associated optical gap. In contrast, one expects a facet-dependent pseudo-relativistic decrease of the optical gap associated with the dispersive part around $\mathbf{k}_\text{line}^D$.
\vspace{1mm}

\begin{figure*}[t]
\begin{center}
\includegraphics[width=.95\textwidth]{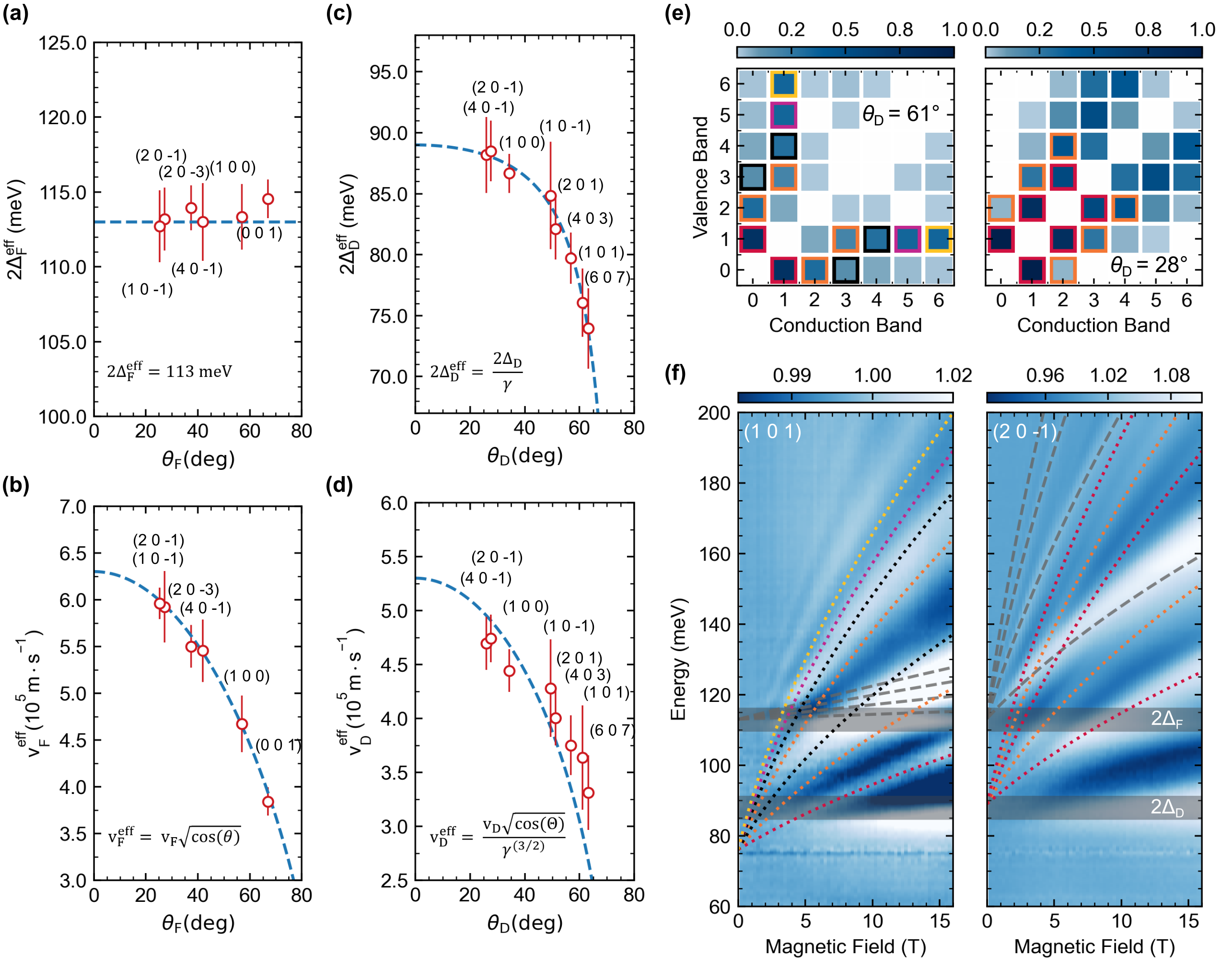}
\end{center}
\caption{ Effective band gap and velocity parameter obtained by a fit of the massive-Dirac model to the lowest inter-LL transition in sets belonging to the flat and dispersive crossings with the Fermi energy: (a),(b) and (c),(d), respectively. (e) Absolute values of matrix elements (the darker color, the stronger the transition) for electric-dipole inter-LL interband excitations (for LLs $n=0\ldots6$ in both conduction and valence bands) calculated for $2\Delta_D=89$~meV, $v_D=5.3\times10^5$~m/s, $w=1.5\times10^5$~m/s and for two different angles $\theta_D=61^{\circ}$ and 28$^{\circ}$ between $\boldsymbol\tau_D$ and $\mathbf{B}$. These two angles correspond to the configuration with $\bf{B}$ perpendicular to the crystallographic planes
(101) and (20$\bar 1$), respectively, for which the experimental $R_B/R_0$ traces are plotted in (f) in a form of false-color plots. The dotted lines show selected inter-LL excitations from the dispersive part of the nodal line, assumed to be electric-dipole active based on the matrix elements presented in (e). We use the same color-framing/coding in (e) and (f) to facilitate the identification of individual transitions. For instance, the lowest (red-dotted) line in (f) corresponds to the $0\leftrightarrow 1$ transitions. The gray dashed lines show the expected transitions in the flat part (selection rules $n\rightarrow n\pm1$, for $v_F=6.3\times10^5$~m/s and $\Delta_F=113$~meV).}
\label{fig:fitting-selectin-rules}
\end{figure*}

These expectations are indeed corroborated by our magneto-optical measurements.
In our data analysis, we associate the maxima in $R_B/R_0$ spectra with positions of inter-LL excitations, a solid assumption in the vicinity of the plasma edge (cf. Figure~\ref{fig:intro-panel}d and~\cite{SM}). The effective values of the band  gap and velocity parameter derived for all explored facets are presented in \textbf{Figure}.~\ref{fig:fitting-selectin-rules}a-d. The response of the flat part matches perfectly the expectations for a Landau-quantized 2D massive Dirac system. The effective band gap $2\Delta_F^{\mathrm{eff}}=(113\pm2)$~meV stays constant within the experimental error (Figure~\ref{fig:fitting-selectin-rules}a).
The variation of the effective velocity (Figure~\ref{fig:fitting-selectin-rules}b)
with the facet reflects the geometrical factor, $v^{\mathrm{eff}}_F=v_F\sqrt{\cos\theta_F}$, where  $v_F=(6.3\pm0.3)\times10^5$~m/s, due to mutual orientation of $\mathbf{B}$ and $\boldsymbol\tau_F$. This allows us to deduce the local direction of the flat part. The best agreement has been found for $\boldsymbol\tau_F$ in the mirror $a$-$c$ plane, $(0\pm2)^{\circ}$, but inclined by $(-23\pm2)^{\circ}$ from the ($a$-$b$) plane, in perfect agreement with conclusions reached by Shao et al.~\cite{ShaoPNAS19}. Notably, for some explored facets, the angle $\theta_F$ approaches $90^{\circ}$ and the upper set of inter-LL excitations can no longer be distinguished in the magneto-reflectivity data (see, e.g.,  Figure~\ref{Fig:color-maps}a,b).
\vspace{1mm}

In contrast, the effective gap $2\Delta^{\mathrm{eff}}_D$ deduced for the dispersive part spreads over an interval greater than 10~meV (Figure~\ref{fig:fitting-selectin-rules}c). To compare this behaviour with our Lorentz-boost-induced gap renormalization (\ref{gapRenorm}),
we have minimized the difference between the theoretical expections and experimental values of the effective gap and velocity, deduced for all facets, by varying parameters $\Delta_D$, $v_D$, $w$ as well as the local direction $\boldsymbol\tau_D$. A very good agreement was obtained (dashed lines in Figure~\ref{fig:fitting-selectin-rules}c,d) for the following parameters: $v_D=(5.3\pm0.5)\times10^5$~m/s, $w=(1.5\pm0.5)\times10^5$~m/s, $2\Delta_D=(89\pm2)$~meV, and the local direction $\boldsymbol\tau_D$ that deviates by $(25\pm10)^{\circ}$ from the mirror ($a$-$c$) plane and by $(-5\pm2)^{\circ}$ from the $a$-$b$ plane. These parameters agree well with the previous estimates given by Shao~et al.~\cite{ShaoPNAS19}. Slightly larger values were found for $w$ and the angle of $\boldsymbol\tau_D$ with respect to the $a$-$c$ plane. Importantly, the angle $\theta_D$ reaches larger values for certain explored facets, thus implying a rapidity that exceeds unity. In such cases, the lower set of inter-LL excitations disappears entirely from the magneto-optical response, suggesting  the complete collapse of the LL spectrum. This happens for the data collected on the (001)-oriented facet ($\theta_D\approx 85^{\circ}$) presented in Figure~\ref{Fig:color-maps}d, and also the (20$\bar{3}$)-oriented facet ($\theta_D\approx 64^{\circ}$), see Supplementary Materials~\cite{SM}.
\vspace{1mm}

In addition to the Lorentz-boost renormalization of the spectrum, our model for the dispersive nodal line implies a departure from the conventional electric-dipole selection rules, $n\rightarrow n\pm1$, which are generally valid for all isotropic systems~\cite{Landwehr2012}.
To illustrate this, we have numerically evaluated
the matrix elements for electric-dipole interband excitations between different pairs of LLs ($n=0\ldots6$) and
visualized them graphically in Figure~\ref{fig:fitting-selectin-rules}e. We have chosen two particular angles, $\theta_D=61^\circ$ and 28$^{\circ}$, which correspond to the magnetic field oriented perpendicular to the (101) and (20$\bar{1}$) planes, respectively. For small angles $\theta_D$, the magneto-optical response
is dominated by $n\rightarrow n\pm1$ transitions, although other excitations emerge as well ({\emph e.g.}, $n\rightarrow n\pm2$). In contrast, for larger angles $\theta_D$, one finds a plethora of optical transitions. The dominant ones
follow the rule-of-thumb selection rules $n\rightarrow \alpha n$ and $n\rightarrow n/\alpha$, where $\alpha$ is an integer ($\alpha = 4-6$ in the left panel of Figure~\ref{fig:fitting-selectin-rules}e), in agreement with preceding works on tilted 3D cones~\cite{SariPRB15,TchoumakovPRL16}. This result may be understood in a broader context of materials which do not have a full rotational symmetry along the direction of the applied magnetic field and in which inter-LL excitations beyond the basic selection rules $n\rightarrow n\pm1$ become electric-dipole active~\cite{OrlitaPRL12,LyJPCM16}.
\vspace{1mm}

As seen in Figure~\ref{fig:fitting-selectin-rules}e, the lowest energy transitions $1\leftrightarrow0$ stay strong as long as $\beta<1$ and were used to deduce
the effective parameters $2\Delta^{\mathrm{eff}}_D$ and $v^{\mathrm{eff}}_D$.
These parameters may now, in turn, be used to identify excitations between LLs with higher indices and thus get experimental insights into the selection rules. To this end, we compare in Figure~\ref{fig:fitting-selectin-rules}f the experimental data collected with $\mathbf{B}$ perpendicular to the (101) and (20$\bar{1}$) crystallographic planes, with the expected positions of selected interband inter-LL excitations (dotted lines) calculated using the corresponding effective gap and velocity parameters. To facilitate the comparison, we use the color-framing/coding introduced in Figure~\ref{fig:fitting-selectin-rules}e.
In line with our expectations, we identify $n\rightarrow n \pm 1$ and $n\rightarrow n \pm 2$ excitations in the response on the (20$\bar{1}$)-oriented facet which implies a relatively small angle $\theta_D$ (Figure~\ref{fig:fitting-selectin-rules}f right). In contrast, when the magnetic field is applied perpendicularly to the (101) crystallographic plane (Figure~\ref{fig:fitting-selectin-rules}f left), we identify transitions with a greater change of the LL index, such as $1\rightarrow 4$ or $1\rightarrow 3$, and no line following the standard $n\rightarrow n\pm1$ selection rule is found, except for the lowest one, $0 \leftrightarrow 1$. In both cases,
the gray dashed lines show the expected response of the flat part of the nodal line that follow the standard $n\rightarrow n \pm 1$ selection rules and no additional excitations emerge, unlike in the dispersive part.

\section{Conclusions}

We have found that the optical band gap of the nodal-line semimetal NbAs$_2$ measured via magneto-optical spectroscopy depends on the facet explored in the experiment. This observation is understood as a consequence of the pseudo-relativistic renormalization of the band gap within a Lorentz boost determined by the slope of the dispersive nodal line. The slope defines, together with the direction of the applied magnetic field, the tilt of the conical dispersion of a massive 2D Dirac electron in the plane perpendicular to the applied magnetic field. Our findings show that the emergent relativistic description of topological quantum materials in terms of Dirac Hamiltonians, or its variants, as well as the use of Lorentz transformations can be pushed surprisingly far. The observed Lorentz-boost-driven renormalization can be also viewed as an analogue of the well-known Franz-Keldysh effect in the magnetic field~\cite{FranzZfN58,AronovJETP67}, nevertheless, in our case, with no real electric field applied.


\section{Experimental Section}
\noindent
\textbf{Sample growth and characterization}

\vspace{0.1mm}
NbAs$_2$ single crystals explored in this work were grown using a chemical vapor transport method. The as-grown crystals usually have a several facets with different crystalographic orientations with shiny surfaces suitable for infrared reflectivity experiments. Individual facets were identified using conventional $x$-ray technique, using a diffractometer equipped with Cu $x$-ray
tube, channel-cut germanium monochromator and scintillation detector.
\vspace{2mm}

\noindent
\textbf{Optical spectroscopy at $B = 0$}

\vspace{0.1mm}
To deduce the optical conductivity of NbAs$_2$ in Figure~\ref{fig:intro-panel}d, the reflectivity on the (001)-oriented facet was measured using radiation polarized linearly along the $a$ and $b$ crystallographic axes. To this end, the Vertex 70v FTIR spectrometer was used, equipped with custom-built in-situ gold-evaporation technique. At high photon energies,
the phase was fixed by ellipsometry. Then, the standard
Kramers-Kronig analysis was employed to obtain the
frequency-dependent complex optical conductivity.
\vspace{2mm}

\noindent
\textbf{Infrared magneto-spectroscopy}

\vspace{0.1mm}
The magneto-reflectivity of NbAs$_2$ was explored in the Faraday configuration, with $\mathbf{B}$ applied perpendicular to the chosen crystallographic plane. During experiments, a macroscopic area of the sample (typically a few mm$^2$), placed in a superconducting coil and kept at $T = 4.2$~K in the helium exchange gas, was exposed to radiation of a globar, which was analyzed by the Vertex 80v FTIR spectrometer and delivered to the sample via light-pipe optics. The reflected light
was detected by a liquid-helium-cooled bolometer placed outside the magnet. The reflectivity $R_B$ recorded at a given magnetic field $B$ was normalized by $R_{B=0}$. In addition, a base-line correction was performed to compensate for variation of the absolute signal intensity over time. To this end, relative magneto-transmission spectra, $R_B/R_0$, were normalized to unity in the spectral range away around $\hbar\omega=400$~meV or above (away from the range of interest).
The reflectivity spectra on each facet were collected using the $\delta B=0.25$~T steps. To create false-color plots of $R_B/R_0$ spectra in Figures~\ref{Fig:color-maps}a-e and \ref{fig:fitting-selectin-rules}f, no linear interpolation was used. Instead, the spectrum collected at the magnetic field of $B$ was plotted in the interval of $B\pm\delta B/2$.
To facilitate the data analysis, we assumed that the maxima in relative magneto-reflectivity, $R_B/R_0$, directly correspond to the positions of inter-LL resonances. This is justified when the imaginary part of the dielectric function exceeds the absolute value of the real part -- a condition fulfilled at photon energies around and slightly above the plasma edge (Figure 1d). A
more detailed analysis indicates that, in this way, we slightly overestimate/underestimate the positions of resonances at lower/higher part of the explored range.
\vspace{2mm}

\medskip
\textbf{Supporting Information} \par 
\vspace{1mm}
Supporting Information is available from the Wiley Online Library or from the author.

\medskip
\textbf{Acknowledgements} \par 
\vspace{1mm}
We acknowledge discussions with T. Brauner and S. Tchoumakov.
The work has been supported by the ANR DIRAC3D project (ANR-17-CE30-0023) and exchange programme PHC ORCHID (47044XE). A. A. acknowledges
funding from the Swiss National Science Foundation
through Project No. PP00P2-170544. M.N.  acknowledges the support of  CeNIKS project co-financed by the Croatian Government and the EU through the European Regional Development Fund - Competitiveness and Cohesion Operational Program (Grant No. KK.01.1.1.02.0013). R.S. acknowledges financial support provided by the Ministry of Science and Technology in Taiwan under project numbers MOST-110-2112-M-001-065-MY3 as well as Academia Sinica
for the budget of AS-iMate-109-13. D.K.M. acknowledges partial support from NSF Grant No. DMR-1914451, and the Research Corporation for Science Advancement through a Cottrell SEED award.
This work was also supported by CNRS through IRP "TeraMIR" and by ANR Colector (ANR-19-CE30-0032).
We also acknowledge the support of the LNCMI-CNRS in Grenoble, a member of the European Magnetic Field Laboratory (EMFL).

\bibliography{Lorentz}

\newpage
\pagenumbering{gobble}

\begin{figure}[htp]
\includegraphics[page=1,trim = 17mm 17mm 17mm 17mm, width=1.0\textwidth,height=1.0\textheight]{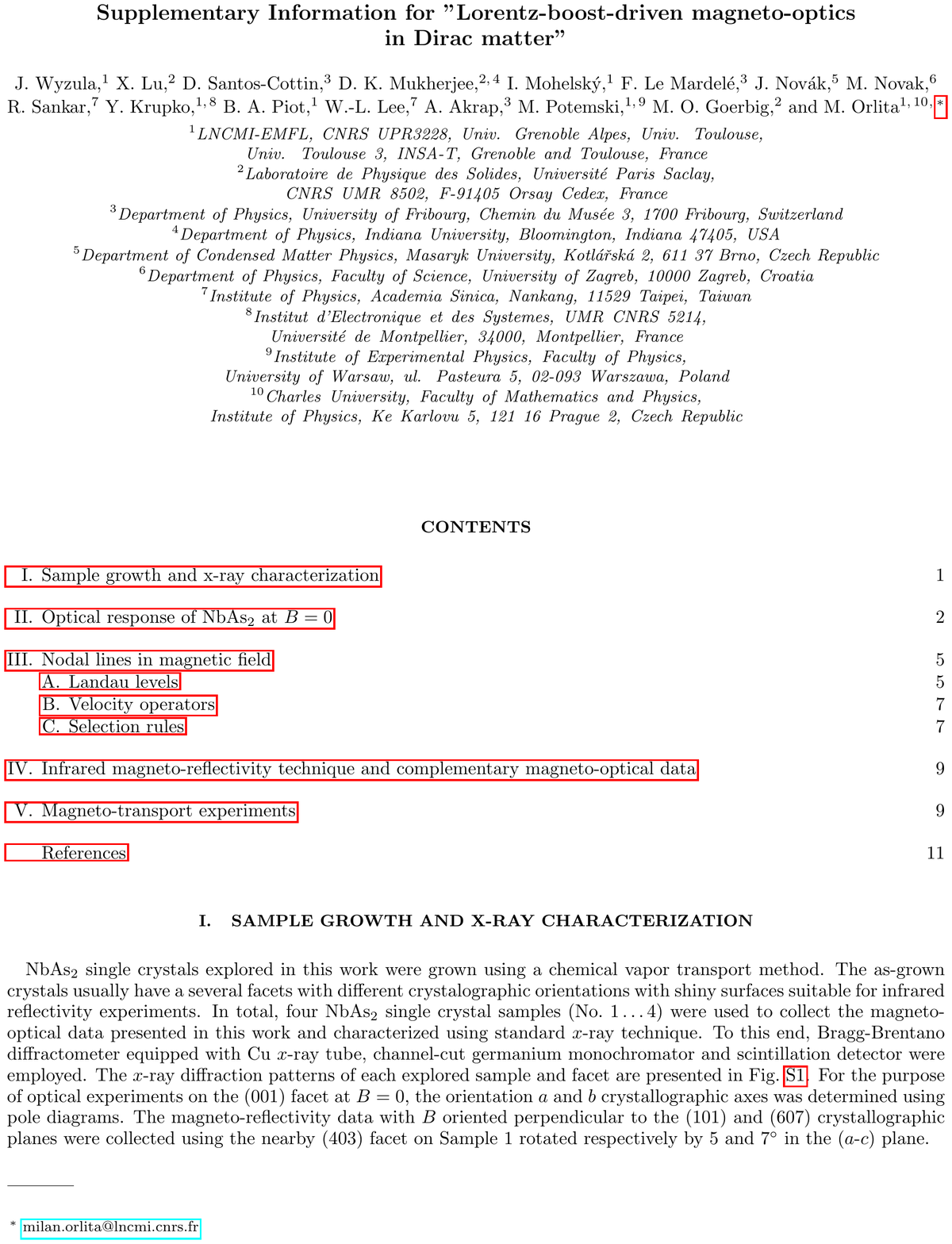}
\end{figure}

\newpage

\begin{figure}[htp]
  \includegraphics[page=2,trim = 17mm 17mm 17mm 17mm, width=1.0\textwidth,height=1.0\textheight]{Supplement.pdf}
\end{figure}

\newpage

\begin{figure}[htp]
  \includegraphics[page=3,trim = 17mm 17mm 17mm 17mm, width=1.0\textwidth,height=1.0\textheight]{Supplement.pdf}
\end{figure}

\newpage

\begin{figure}[htp]
  \includegraphics[page=4,trim = 17mm 17mm 17mm 17mm, width=1.0\textwidth,height=1.0\textheight]{Supplement.pdf}
\end{figure}

\newpage

\begin{figure}[htp]
  \includegraphics[page=5,trim = 17mm 17mm 17mm 17mm, width=1.0\textwidth,height=1.0\textheight]{Supplement.pdf}
\end{figure}

\newpage

\begin{figure}[htp]
  \includegraphics[page=6,trim = 17mm 17mm 17mm 17mm, width=1.0\textwidth,height=1.0\textheight]{Supplement.pdf}
\end{figure}

\newpage

\begin{figure}[htp]
  \includegraphics[page=7,trim = 17mm 17mm 17mm 17mm, width=1.0\textwidth,height=1.0\textheight]{Supplement.pdf}
\end{figure}

\newpage

\begin{figure}[htp]
  \includegraphics[page=8,trim = 17mm 17mm 17mm 17mm, width=1.0\textwidth,height=1.0\textheight]{Supplement.pdf}
\end{figure}

\newpage

\begin{figure}[htp]
  \includegraphics[page=9,trim = 17mm 17mm 17mm 17mm, width=1.0\textwidth,height=1.0\textheight]{Supplement.pdf}
\end{figure}

\newpage

\begin{figure}[htp]
  \includegraphics[page=10,trim = 17mm 17mm 17mm 17mm, width=1.0\textwidth,height=1.0\textheight]{Supplement.pdf}
\end{figure}

\newpage

\begin{figure}[htp]
  \includegraphics[page=11,trim = 17mm 17mm 17mm 17mm, width=1.0\textwidth,height=1.0\textheight]{Supplement.pdf}
\end{figure}

\begin{figure}[htp]
  \includegraphics[page=12,trim = 17mm 17mm 17mm 17mm, width=1.0\textwidth,height=1.0\textheight]{Supplement.pdf}
\end{figure}

\end{document}